# Multifragmentation and nuclear phase transitions (liquid-fog and liquid-gas)

V.A. Karnaukhov[a], H. Oeschler[b], S.P. Avdeyev[a], V.K. Rodionov[a], A.V. Simomenko[a],
V.V. Kirakosyan[a], A. Budzanowski[c], W. Karcz[c], I. Skwirczynska[c], E.A. Kuzmin[d],
E. Norbeck[e], A.S. Botvina[f]

[a] Joint Institute for Nuclear Research, Dubna  (karna@nusun.jinr.ru)

[b] Institut für Kernphysik, Darmstadt University of Technology, Darmstadt,  Germany

[c] H.Niewodniczanski Institute of Nuclear Physics, Cracow, Poland

[d] Kurchatov Institute,  Moscow

[e] University of Iowa, Iowa City, USA

[f] GSI, Darmstadt, Germany

   Thermal multifragmentation of hot nuclei is interpreted as the nuclear *liquid-fog* phase
transition.  The charge distributions of the intermediate mass fragments produced      in
$p(3.6$ GeV$) +$ Au and $p(8.1$ GeV$) +$ Au collisions are analyzed within the statistical
multifragmentation model with the critical temperature for the *nuclear liquid-gas* phase
transition $T_c$ as a free parameter. The analysis presented here provides strong support for a
value of $T_c > 15$ MeV.

## 1. INTRODUCTION

   The investigation of the decay properties of very hot nuclei is currently one of the most
challenging topics of nuclear physics. The excitation energy of the hot nuclei is comparable
with the total binding energy. They disintegrate via a new multibody decay mode–*thermal
multifragmentation* [1], which is characterized by the copious emission of intermediate mass
fragments ($2 < Z \leq 20$). The development of this field has been strongly stimulated by an idea
that this process is related to the nuclear liquid-gas phase transition, which was predicted on
the basis of the similarity between van der Waals and nucleon-nucleon interactions [2-4]. The
equations of state for the two cases are similar. For both systems there is a spinodal region
characterized by phase instability. The density here is reduced as compared to the liquid
phase. One can imagine that a hot nucleus (at $T = 5$-7 MeV) expands due to thermal pressure
and enters the unstable region [5]. Due to density fluctuations, a homogeneous system
converts into a mixed phase state consisting of droplets (IMF) surrounded by nuclear gas. The
final state of this transition is a *nuclear fog* [4], which explodes due to Coulomb repulsion and
is detected as multifragmentation. The disintegration time is very short. This is the scenario of
a spinodal decomposition. It was proven experimentally that thermal multifragmentation
occurs at reduced (3-4 times) densities, and the disintegration time is less than 100 fm/$c$. The
spinodal decomposition is, in fact, the *liquid-fog* phase transition in a nuclear system.

An important model parameter of this scenario is the critical temperature $T_c$ for the *nuclear liquid-gas* phase transition. The surface tension vanishes at $T_c$, and only the gas phase is possible above this temperature. There are many calculations of $T_c$ for finite nuclei. In Refs. [2,3,6,7] it is done by using a Skyrme interaction and the thermal Hartree-Fock theory. The values of $T_c$ were found to be in the range 10-20 MeV depending on the details of the model. The main source of the experimental information for $T_c$ is the fragment yield. In statistical models of nuclear multifragmentation the shape of the IMF charge distribution $Y(Z)$ is sensitive to the ratio $T/T_c$. The charge distribution is well described by the power law $Y(Z) \sim Z^{-\tau}$ for a wide range of colliding systems. In earlier studies the power-law behavior of the IMF yield was interpreted as an indication of the proximity of the excited system to $T_c$. This was stimulated by the application of Fisher's classical droplet model [8], which predicts a pure power-law droplet-size distribution with $\tau = 2$-3 at the critical point. A more sophisticated use of this model for the estimation of $T_c$ has been made in Refs. [9,10]. The data obtained for $\pi$ (8 GeV/c) + Au collisions were analyzed giving $T_c$= (6.7±0.2) MeV. The same analysis was applied to the data for collisions of Au, La, Kr (at A·1.0 GeV) with carbon. The extracted values of $T_c$ were (7.6±0.2), (7.8±0.2) and (8.1±0.2) MeV respectively.

Having in mind the shortcomings of Fisher's model [11,12], we have estimated the nuclear critical temperature using the statistical multifragmentation model (SMM) [13] as it describes well the different experimental data. It was found that $T_c$=20±3 MeV [14]. Some additional analysis is given below with emphasis put on the model dependence of the results.

## 2. ESTIMATION OF $T_C$ USING SMM

Within this model one considers a microcanonical ensemble of all break-up channels composed of nucleons and excited fragments. It is assumed that an excited nucleus expands and then breaks up into nucleons and hot fragments. It is also assumed that at the break-up the nucleus is in thermal equilibrium characterized by the channel temperature $T$. The charge yield depends on the contribution of the surface free energy of fragments $F_{AZ}^S$ to the entropy of a given final state. The surface energy depends on the critical temperature:

$$F_{AZ}^S = a_s(T)A^{2/3}, a_s(T) = a_s(0)\left(\frac{T_c^2 - T^2}{T_c^2 + T^2}\right)^{5/4}, \qquad (1)$$

with $a_s(T) = 4\pi r_0^2 \sigma(T)$, where $\sigma(T)$ is a surface tension coefficient. This equation was obtained in [15]. It is successfully used by the SMM for describing the decay of hot finite nuclei. The comparison of the measured and calculated IMF charge yields is the way to estimate the $T_c$ value.

The reaction mechanism is usually divided into two stages. The first one is a fast energy-depositing stage. We use the intranuclear cascade model (INC) [16] for describing this stage. The second stage is described by the SMM, which considers decay of a hot target spectator. But such an approach fails to explain the IMF multiplicities. An expansion stage (Exp.) is inserted between the two parts of the calculation. The excitation energies and the residual masses are then fine tuned [1,5] to get agreement with the measured IMF multiplicities. Figure 1a presents the measured fragment charge distribution for $p$(8.1 GeV)+Au collisions and the calculations performed with $T_c$ as a free parameter. Experiments were done using the

4π-setup FASA [5]. The lines show the calculated distributions for $T_c$ = 7, 11 and 18 MeV. The SMM parameter $k$ is set to 2. It corresponds to the break-density $\rho_f = 1/3\rho_o$. The level density parameter is taken to be $a = A/8$. The statistical errors of the data do not exceed the size of the dots. The calculations are close to the data for $T_c$ = 18 MeV. The estimated mean temperature of the system is around 6 MeV, the mean charge and mass numbers are 67 and 158. The theoretical curves deviate from the data with decreasing $T_c$. Figure1b gives the results of the power-law fits for the data and model calculations. The Be yield was corrected in the fitting procedure for the loss of unstable $^8$Be. Similar results are obtained for $p$+Au collisions at 3.6 GeV. Comparisons of the experimental power-law exponents and model predicted ones for different values of $T_c$ are shown in Fig.2. In contrast to our paper [14] the power-law fits are done in the range Z=4-11 to exclude the influence of nonequibrium emission of Li [1]. The measured power-law exponents are given as a band with a width determined by the statistical error. The size of the symbols for the calculated values of $\tau_{app}$ is of the order of the error bar. From the best fit of the data for both beam energies and the calculations the critical temperature is (17±2) MeV. Calculations were also performed with the surface tension coefficient linearly dependent on $T/T_c$ as in [9,10] (middle panel, solid circles). The predicted values of $\tau_{app}$ in this case are remarkably lower than the data. The results only slightly depend on the level density parameter. This is seen from the bottom panel where upper line is calculated with

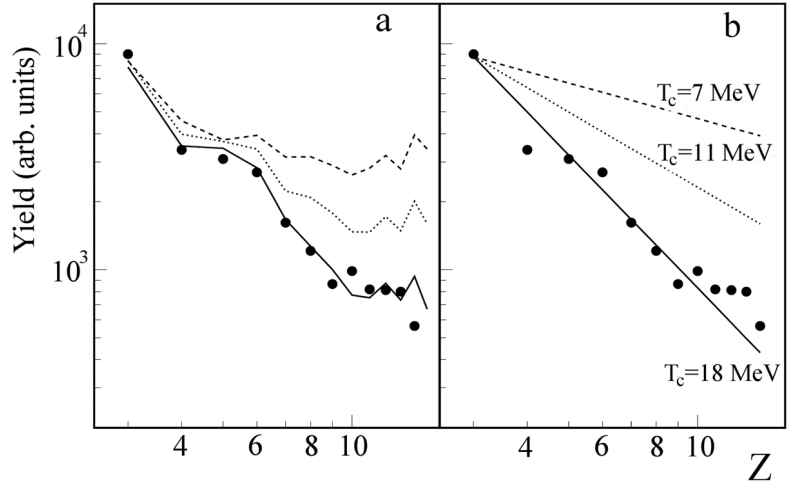

**Fig. 1.** IMF charge distributions for for $p$ + Au at 8.1 GeV: data and the INC+Exp+SMM model predictions (a), power law fits (b).

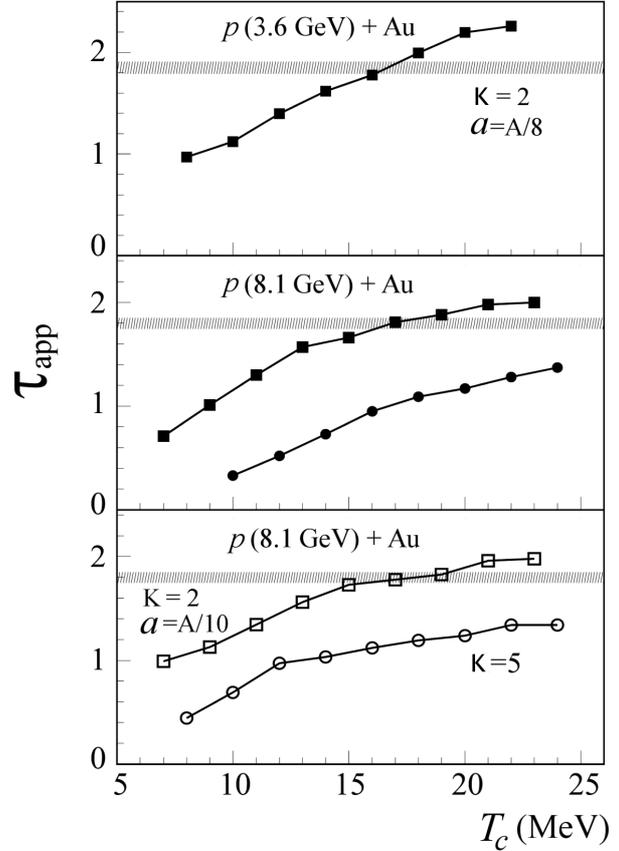

**Fig.2.** Power-law exponents for the IMF charge distributions for $p$+Au collisions.

$a$ = A/10. The lower line in Fig.2 is obtained on the assumption of $k$ =5, that is $\rho_f$= 1/6 $\rho_o$. The model prediction does not agree with the data. Note that our choice of $k$ = 2 is motivated by analysis of the shape of the IMF energy spectra, which is sensitive to the size of the fragmenting nucleus [5].

## 3. SUMMARY

Thermal multifragmentation of hot nuclei is interpreted as the *liquid-fog* phase transition. The critical temperature for the nuclear *liquid-gas* phase transition $T_c$ (at which the surface tension vanishes) is estimated by using the statistical multifragmentation model. For that purpose, the IMF charge distributions for $p$ + Au collisions at 3.6 and 8.1 GeV have been analyzed within the SMM with $T_c$ as a free parameter. The value $T_c$ = (17 ± 2) MeV obtained from the best fit to the data should be considered as some effective value of $T_c$ averaged over all the fragments produced in the collision. This value is larger than those found in [9,10] using Fisher's droplet formalism, but it is close to the values obtained in [17,18]. Although our value for $T_c$ is model dependent, as is any other estimate of the critical temperature, the analysis presented here provides strong support for a value of $T_c$ > 15 MeV.

The research was supported in part by Grant 03-02-17263 from the Russian Foundation for Basic Research, by the Grant of the Polish Plenipotentiary to JINR, and by Contract 06DA453 with BFT(Germany).

## RFFERENCES

1. S.P. Avdeyev, et al., Nucl. Phys. A 709 (2002) 392.
2. G. Sauer, H. Chandra, U. Mosel, Nucl. Phys. A 264 (1976) 221.
3. H. Jaqaman, A.Z. Mekjian, L. Zamick, Phys. Rev. C 27 (1983) 2782.
4. P.J. Siemens, Nature 305 (1983) 410; Nucl. Phys. A 428 (1984) 189c.
5. S.P. Avdeyev, et al., Eur. Phys. J. A 3 (1998) 75.
6. P. Bonche, et al., Nucl. Phys. A 436 (1985) 265.
7. Zhang Feng-Shou, Z. Phys. A 356 (1996) 163.
8. M.E. Fisher, Physics. 3 (1967) 255.
9. J.B. Elliott, et al., Phys. Rev. Lett. 88 (2002) 042701.
10. J.B. Elliott, et al., nucl-ex /0205004v1 (2002).
11. J. Schmelzer, G. Ropke, F.P. Ludwig, Phys. Rev. C 55 (1997) 1917.
12. P.T. Reuter, K.A. Bugaev, Phys. Lett. B 517 (2001) 233.
13. J. Bondorf, et al., Phys. Rep. 257 (1995) 133.
14. V.A. Karnaukhov, et al., Phys. Rev. C 67 (2003) 011601(R).
15. D.G. Ravenhall, C.J. Pethick, J.M. Lattimer, Nucl. Phys. A 407 (1983) 571.
16. V.D. Toneev, et al., Nucl. Phys. A 519 (1990) 463c.
17. J.B. Natowitz, et al., Phys. Rev. Lett. 89 (2002) 212701.
18. R. Ogul, A.S. Botvina, Phys. Rev. C 66 (2002) 051601(R).